\documentclass[twocolumn]{emulateapj}

\usepackage{graphicx}
\usepackage{epsfig}
\usepackage{amsmath}

\usepackage{color}

\shorttitle{Effects of Spot Size on NS Pulse Profiles}
\shortauthors{Baub\"ock et al.}

\begin{document}


\title{Effects of Spot Size on Neutron-Star Radius Measurements from Pulse Profiles}

\author{Michi Baub\"ock, Dimitrios Psaltis, and Feryal \"Ozel}
\affil{Astronomy Department,
University of Arizona,
933 N.\ Cherry Ave.,
Tucson, AZ 85721, USA}
\email{mbaubock@email.arizona.edu}

\begin{abstract}
We calculate the effects of spot size on pulse profiles of moderately
rotating neutron stars. Specifically, we quantify the bias introduced
in radius measurements from the common assumption that spots are
infinitesimally small. We find that this assumption is reasonable for
spots smaller than 10$^\circ$--18$^\circ$ and leads to errors that are
$\le$10\% in the radius measurement, depending on the location of the
spot and the inclination of the observer. We consider the implications
of our results for neutron star radius measurements with the upcoming
and planned X-ray missions NICER and LOFT. We calculate the expected
spot size for different classes of sources and investigate the
circumstances under which the assumption of a small spot is justified.
\end{abstract}


\section{Introduction}
Neutron stars provide a natural testbed for a variety of interesting
physical phenomena. At their cores, neutron stars contain matter at
densities and pressures unachievable in the laboratory setting. By
measuring their properties, it is possible to constrain the unknown
equation of state of cold dense matter. In particular, simultaneous
mass and radius measurements with better than ten percent accuracy are
needed to reach this goal (e.g., Lindblom 1992; Lattimer \& Prakash
2001; Read et al.\ 2009; \"Ozel \& Psaltis 2009). Moreover, such
measurements can further be used for testing the limits of General
Relativity and alternate theories of gravity (e.g., Psaltis 2008).

One promising avenue for simultaneous mass and radius measurements is
through the observations of pulse profiles originating from
temperature anisotropies on the stellar surface. These anisotropies
appear on a variety of isolated and accreting sources. In isolated
pulsars, magnetic return currents heat the region around the magnetic
pole, creating a local hotspot (e.g., Ruderman \& Sutherland 1975;
Arons 1981; Harding \& Muslimov 2001, 2002; Bai \& Spitkovsky 2010a,
b). In accreting neutron stars, the hotspots arise from matter
funneled onto a small region of the surface (see Frank et al.\ 2002
and references therein). Alternatively, in X-ray bursters, the
thermonuclear burning is thought to ignite at one spatially localized point in the
accreted layer and takes many spin periods to spread across the
stellar surface, creating a short-lived temperature anisotropy (e.g.,
Schoelkopf \& Kelley 1991; Strohmayer 1992, Strohmayer et al.\ 1996; Bildsten
1995). In all of these cases, the amplitude and shape of the pulse
profile encodes information about the size and location of the spot as
well as the properties of the neutron star.

\begin{figure*}
\psfig{file=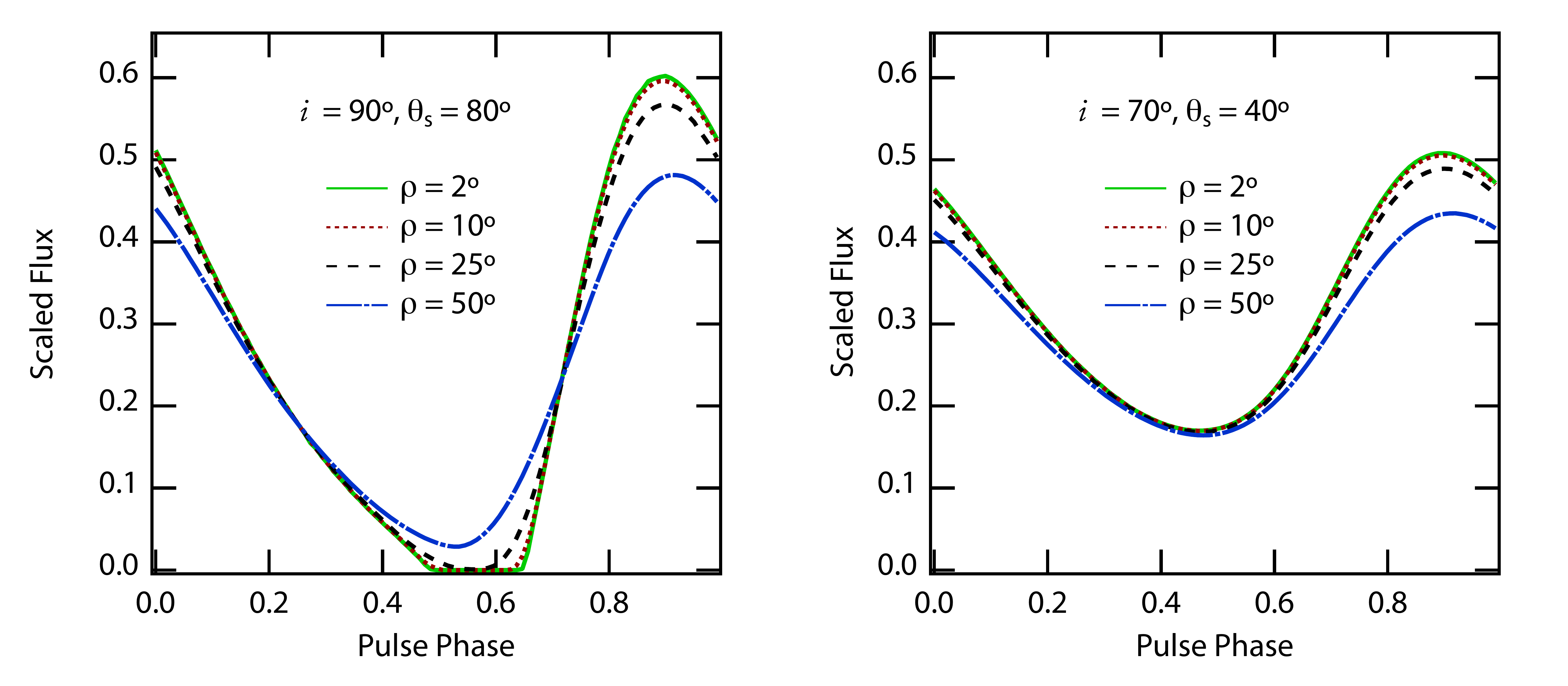, width=7in}
\caption{Example pulse profiles for two different spot
  configurations. In the left panel, the observer inclination is set
  to $i = 90^\circ$ and the spot colatitude to $\theta_s =
  80^\circ$. In the right panel, they are set to $i = 70^\circ$ and
  $\theta_s = 40^\circ$. In both panels, the angular radius of the
  spot varies from $2^\circ$ to $50^\circ$. The flux has been scaled
  by $(50^\circ/\rho)^2$ in order to remove the primary dependence of
  the flux on the spot area. In all cases, we have set $M =
  1.6$~M$_\odot$, $R = 10$~km, and $f_{\rm NS} = 600$~Hz.}
\label{fig:Example_Pulses}
\end{figure*}

In order to extract the mass and radius from the pulse profile shape,
accurate theoretical models are needed that take account gravitational
lensing, Doppler shifts, and time delays from the neutron-star
surface. Several studies to date have focused on modeling pulse
profiles, including a variety of relativistic effects. The simplest
approximation is to model the star as a sphere in the Schwarzschild
metric and separately calculate the Doppler shift and aberration due
to the rotation of the stellar surface (e.g., Miller \& Lamb 1998;
Muno et al.\ 2002, 2003; Lamb et al.\ 2009a, b; Lo et
al.\ 2013). Poutanen \& Beloborodov (2006, see also Bogdanov et
al.\ 2007) used this model to find analytic approximations for the
photon lensing and time delays and constructed analytic lightcurves
for neutron-star pulse profiles.

The Schwarzschild metric describes a spherical non-spinning mass and
is, therefore, an appropriate limit for slowly spinning neutron
stars. For faster spins, however, effects at increasing order of spin
frequency become important. At first order in spin frequency, frame
dragging affects the spacetime around a spinning star, which is
described by the Kerr metric. Braje et al.\ (2000) modeled pulse 
profiles in the Kerr metric and found that the distortion of the
profile due to these effects is at the 1\% level. At second order in
spin frequency, the neutron star becomes oblate in shape and acquires
a significant quadrupole moment. Cadeau et al.\ (2007, see also
Morsink et al.\ 2007) found that the oblateness significantly alters
the pulse profile, causing the spot to be visible at inclinations
where it would be eclipsed by a spherical star. Psaltis \& \"Ozel
(2014) additionally included the effect of an appropriate quadrupole
moment by using the Hartle-Thorne metric, which is formally correct up
to second order in spin frequency. Even higher orders in spin
frequency can introduce additional effects to the spacetime. Cadeau et
al.\ (2007) modeled lightcurves for rapidly spinning neutron stars by
calculating the metric numerically. They found that corrections from
higher-order terms are negligible for the spin frequencies of known
neutron stars.

The problem of modeling pulse profiles and, in turn, inferring
neutron-star properties from pulse profiles, is difficult because of
the large number of parameters needed. In principle, for any
neutron-star model, the pulse profile depends on the properties of the
neutron star as well as the size, shape, and position of the hotspot
on the stellar surface. In order to simplify the models and reduce the
parameter space, pulse profiles are often calculated under the
assumption that the spot is infinitesimal in size. Lamb et
al.\ (2009a, b; see also Psaltis et al.\ 2000) considered the effects
of spot size on the amplitude of the pulse profile and found that the
spot size has negligible effect as long as it is below
$\sim45^\circ$. However, they did not consider the effect of the spot
size on the higher harmonics of the profile, which are necessary for
measuring the stellar radius.

In this paper, we investigate the effect of spot size on pulse
profiles for moderately spinning sources. Specifically, we determine
the maximum spot size for which the spot can be considered small for
the purpose of measuring the neutron-star radius. We then estimate the
expected spot sizes as a function of spin frequency for
rotation-powered and accretion-powered X-ray pulsars. In the case of
X-ray burst oscillations, we find the fraction of the burst rise times
during which a sufficient number of counts can be accumulated before
the spot size becomes large enough to significantly bias the radius
measurement from the pulse profile. We discuss the implications of our
results for future missions that will observe pulse profiles, such as
NICER and LOFT.

\section{Pulse Profiles}

We use the ray-tracing code described in Baub\"ock et al.\ (2012) and
Psaltis \& \"Ozel (2014), which calculates pulse profiles using the
Hartle-Thorne metric to approximate the spacetime around a moderately
spinning neutron star. This metric is formally correct to second order
in spin frequency and allows for the stellar surface to become oblate
and the mass distribution to acquire an appropriate quadrupole
moment. Our algorithm also accounts for frame dragging around the
star, time of flight delays, and Doppler shifts and aberration due to
the motion of the stellar surface.

The propagation of photons to infinity depends on seven parameters of
the neutron star: the mass, the equatorial radius, the spin frequency,
the moment of inertia, the oblateness, the quadrupole moment, and the
inclination of the observer's line of sight to the spin axis. Of
these, the mass, spin frequency, and inclination are independent and
vary between different sources. The radius, oblateness, moment of
inertia, and quadrupole moment of the star are determined, for a given spin, by the
equation of state of the matter in its interior. However, there exist
empirical relations between several of these parameters that hold to
high accuracy, for a range of equations of state (e.g., Baub\"ock et
al.\ 2013; Yagi \& Yunes 2013). We use the relations in Baub\"ock et
al.\ (2013) to calculate the oblateness, moment of inertia, and
quadrupole moment for a star with a given mass, radius, and spin
frequency.

By using the universal relations described above, we reduce the number
of parameters for each pulse profile to six: the mass and radius of
the neutron star, the spin frequency, the inclination of the observer
to the spin axis, and the colatitude and angular radius of the
hotspot. For the simulations presented in this work, we fix the mass
at 1.6~M$_\odot$, the radius at 10~km, and the spin frequency at
600~Hz. We calculate pulse profiles over the full range of observer
inclinations $i$ and spot colatitudes $\theta_s$. We define the
observer's inclination as the angle between the the observer's line of
sight and the spin axis, such that $i = 0^\circ$ and $i = 180^\circ$
correspond to an observer directly over the north and south pole of
the neutron star, respectively, while $i = 90^\circ$ corresponds to an
observer in the equatorial plane. We vary the colatitude of the spot
$\theta_s$, such that a spot on the pole has a colatitude $\theta_s =
0^\circ$ and a spot on the equator corresponds to $\theta_s =
90^\circ$.

Figure~\ref{fig:Example_Pulses} shows several example pulse profiles
for a variety of parameters. For all neutron-star configurations, the
flux observed from the hotspot is proportional (to first order) to its
area. In this figure, we have scaled the flux by this approximation to
the spot area in order to highlight the more subtle changes in the
profile shape introduced by increasing the spot size. In the left
panel, the observer is in the equatorial plane of the neutron star,
while the spot is near the equator. In this configuration, the spot is
eclipsed behind the neutron star for a fraction of the spin period. As
the spot grows in size, the duration of the eclipse decreases until
the spot is visible at all phases.

The right panel of Figure~\ref{fig:Example_Pulses} shows profiles for
a smaller observer inclination and spot colatitude. In this case, the
spot is visible at all phases, regardless of its size. At larger
angular radii, however, the amplitude of the profile still
decreases. Moreover, the higher order harmonics also decrease, leading
to a more symmetric and sinusoidal profile.

Since observations of neutron-star pulse profiles are usually limited
by photon counts, information is often extracted by means of
decomposing profiles into their Fourier components (Poutanen \&
Beloborodov 2006; Psaltis et al.\ 2014). The number of parameters of
the neutron star and the hotspot that can be constrained increases
with the number of Fourier components that can be measured from the
lightcurve. We limit our analysis to realistic situations where the
fundamental and the second harmonics can be measured in at least two
energy bands.

 \begin{figure}
\psfig{file=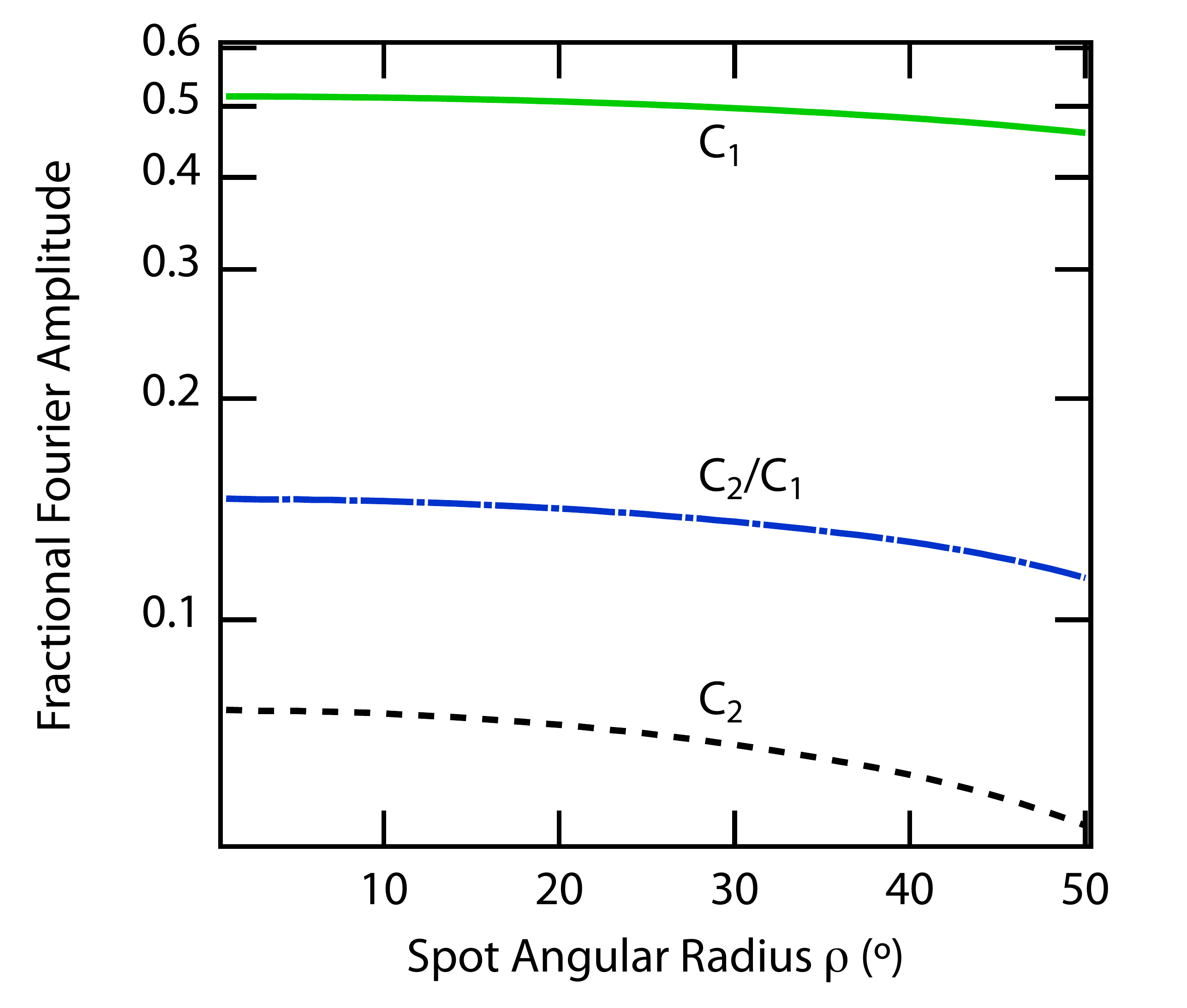, width=3.5in}
\caption{Fractional Fourier amplitudes of pulse profiles as a function
  of spot size. The observer inclination is $i = 70^\circ$ and the
  spot colatitude is $\theta_s = 40^\circ$. The remaining parameters
  are as in Figure~\ref{fig:Example_Pulses}. The solid and dotted
  lines show the first and second Fourier amplitudes,
  respectively. The dashed-dotted line shows the ratio between the two
  components.}
\label{fig:Fourier_Amplitudes}
\end{figure} 
 
As expected from the examples in Figure~\ref{fig:Example_Pulses}, in
addition to encoding information about the neutron star itself and the
location of the hotspot, the Fourier amplitudes of the pulse profiles
also depend on the size of the hotspot. In particular, as the size of
the spot increases, the higher order harmonics are damped and the
profile becomes more sinusoidal. In
Figure~\ref{fig:Fourier_Amplitudes}, we show the amplitude of the
first and second Fourier components, labeled $C_1$ and $C_2$
respectively, for a specific configuration of the observer inclination
and the spot colatitude. For larger spot sizes, both the first and
second harmonics decrease in amplitude. However, the second harmonic
decreases more quickly than the first, causing the ratio between them
to decrease also, as shown by the dashed-dotted line. This introduces a
potential bias in the measurement of neutron-star radii, since the
principal effect of a smaller radius is the reduction of the
higher-order harmonics and, thereby, of the ratio $C_2/C_1$.

\section{Biases in Radius Measurements}

We wish to determine the extent to which the common assumption of an
infinitesimally small spot is justified for the purpose of measuring
the radius of a neutron star from its pulse profile. In order to
quantify the effect of the spot size on measurements of the radius, we
calculate profiles with a very small spot (with an angular radius
$\rho = 2^\circ$) and quantitatively compare these to profiles with
larger spots. We find the maximum spot size for which the assumption
that the spot is infinitesimally small is reasonable, i.e., leading to
a radius bias that is less than 10\%, allowing for meaningful
constraints on the equation of state.

Using the relationships from Baub\"ock et al.\ (2013) for a neutron
star of known spin frequency and the assumption that the hotspot is
small, one can reduce the number of model parameters that need to be
determined from the pulse profiles to four: the mass and radius of the
neutron star, the colatitude of the hotspot, and the inclination of
the observer. The number of observable quantities obtained from a
pulse profile, on the other hand, depends to some extent on the
relative inclinations of the observer and the spot to the spin axis of
the neutron star. We will consider two cases: when the spot is visible
for all phases, and when the spot becomes eclipsed behind the star for
part of the spin period.

In the first case, we can measure the radius directly using the first
two Fourier components of the pulse profiles at two broad energy
ranges as the four required observables (see discussion in Psaltis et
al.\ 2014). All four observables carry information about the radius of
the neutron star because Doppler shifts and aberration, which are
proportional to the stellar radius, affect their values. Following
Psaltis et al.\ (2014), we will use the ratio of amplitudes of the
first two Fourier harmonics, which to leading order is proportional to
the Doppler shift (see below), in order to quantify the effect of the
spot size on the radius measurement.

The second case arises when both the hotspot and the observer are
located near the stellar equator. Although the strong lensing of
photons increases the fraction of the surface area visible to a
distant observer, there is a region which is hidden from sight in this
configuration. When the spot is in this region, none of its flux
reaches the observer. This eclipse introduces sharp edges to the pulse
profile and causes many higher harmonics to become large. However, the
amplitudes of these higher order harmonics are not independent, as
they are caused by the sharp eclipse. In practice, the eclipse
introduces one additional measurable parameter, which is its
duration. For a given spot colatitude and observer's inclination, a
spot with a smaller radius will tend to have shorter eclipses, as the
lensing is stronger and a smaller fraction of the stellar surface is
hidden. However, a large hotspot can mimic the effects of a smaller
radius. As the spot grows, the fraction of the profile during which it
is completely hidden from the observer decreases. In this second case,
we will determine the effect of the spot size on the eclipse duration
in order to quantify the possible bias in radius measurements.

In order to estimate the error introduced in the radius measurement by
a non-negligible spot size, we use the relation between the radius and
the ratio of the first and second harmonics (Psaltis et al.\ 2014)
\begin{equation}
\frac{C_2}{C_1} \approx \frac{4 \pi f}{c} R \sin{i}\sin{\theta_s}.
\label{eq:R_Cratio}
\end{equation}
This scaling was derived for a slowly spinning star in the
Schwarzschild metric and is only expected to hold approximately for
the most rapidly spinning stars. However, our purpose here is only to
approximate the error in the radius measurement introduced by large
spot sizes, for which this expression is sufficient.

Our goal is to calculate the maximum spot size at which the change
introduced to the ratio $C_2/C_1$ due to the spot size is comparable
to the difference introduced to this quantity by changing the radius
by 10\%. To determine this maximum size, we first calculate the
change in the ratio $C_2/C_1$ as
\begin{equation}
\Delta \left(\frac{C_2}{C_1}\right) \approx 0.025 \left(\frac{\Delta R}{\rm 1~km}\right) 
\left(\frac{f}{\rm 600~Hz}\right) \sin{i}\sin{\theta_s}.
\label{eq:Delta_Cratio}
\end{equation}
Next, we calculate the change in this ratio introduced by increasing
the spot size. Specifically, for each pair of angles $i$ and
$\theta_s$, we calculate a pulse profile and its Fourier components
for a range of spot sizes between $2^\circ$ and $50^\circ$. As the
spot grows in size, there is an increasing change in the
ratio $C_2/C_1$. Comparing the two, we find the largest spot size
$\rho$ for which this change is smaller than the error corresponding
to a radius uncertainty of 10\% given by
equation~(\ref{eq:Delta_Cratio}).

\begin{figure}
\psfig{file=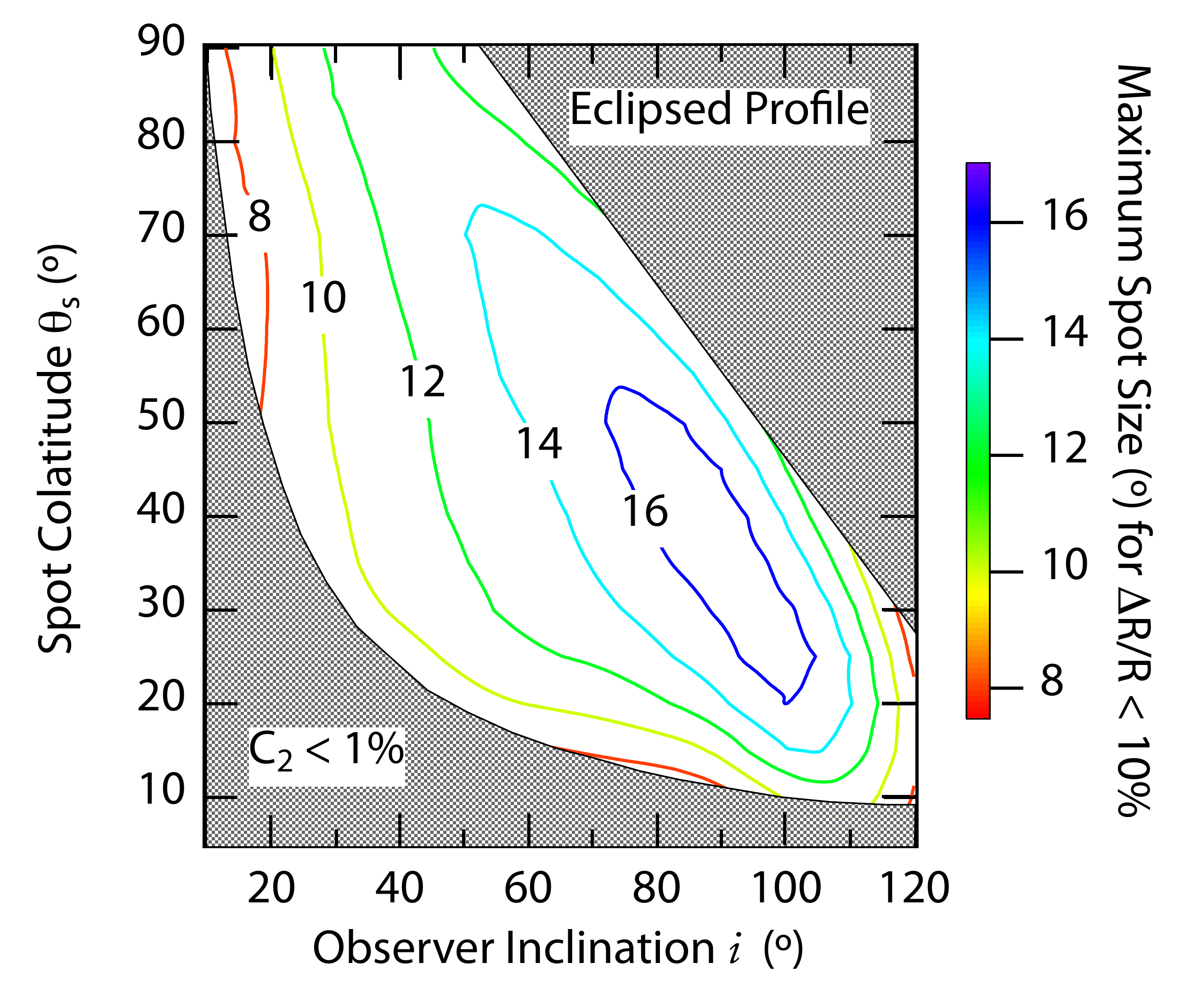, width=3.5in}
\caption{Maximum spot angular radius that leads to a $\le$10\% error
  in radius measurement as a function of the spot colatitude and
  observer inclination. If a radius measurement is made under the
  assumption of an infinitesimally small spot, the contours correspond
  to the spot size for which the measurement differs from the true
  neutron-star radius by ten percent. }
\label{fig:Max_Spot_Size}
\end{figure}

The contours in Figure~\ref{fig:Max_Spot_Size} show the spot size
which corresponds to a maximum error of 10\% in radius. For most of
the parameter space, we find that a spot size of up to 10$^\circ$--18$^\circ$ introduces biases in the radius that are smaller than
10\%. The maximum allowable size generally increases with increasing
inclination and decreasing spot colatitude. This effect is primarily
due to the $\sin i \sin \theta_s$ term in
equation~(\ref{eq:Delta_Cratio}). For small angles $i$ and $\theta_s$,
the change in the ratio $C_2/C_1$ for a 10\% change in the
neutron-star radius becomes very small, leading to a small allowed
spot size.

The contours in Figure~\ref{fig:Max_Spot_Size} are calculated for a 10~km star. For stars with larger radii, the maximum allowed spot size that leads to a 10\% bias in the radius measurement is smaller. This is due to the larger tangential velocity of the neutron star surface for a given spin frequency. As the linear velocity increases, the differential Doppler boost from the near and far edges of the spot reduces the amplitude of the second harmonic of the pulse profile more than for a smaller star at the same spin frequency. This leads to a larger bias in the radius measurement and therefore a smaller maximum spot size. 

The Doppler boost scales as the tangential velocity squared and thus as $R^2$. Therefore, the spot size that introduces a fixed absolute bias in the radius measurement scales to first order as $1/R^2$. In order to achieve a fractional accuracy of 10\%, the maximum spot size scales as $1/R$.

The shaded parts of the parameter space in Figure~\ref{fig:Max_Spot_Size} correspond to regions where
our calculation of the maximum spot size is no longer valid or the
measurement becomes unfeasible. In the lower region, corresponding to
configurations where the observer and the spot are both near the
rotational pole of the neutron star, the amplitude of the pulse
profile is very small and the second harmonic is less than one
percent. In the shaded region, the small amplitudes of pulse profiles
make radius measurements of neutron stars unfeasible.

\begin{figure}
\psfig{file=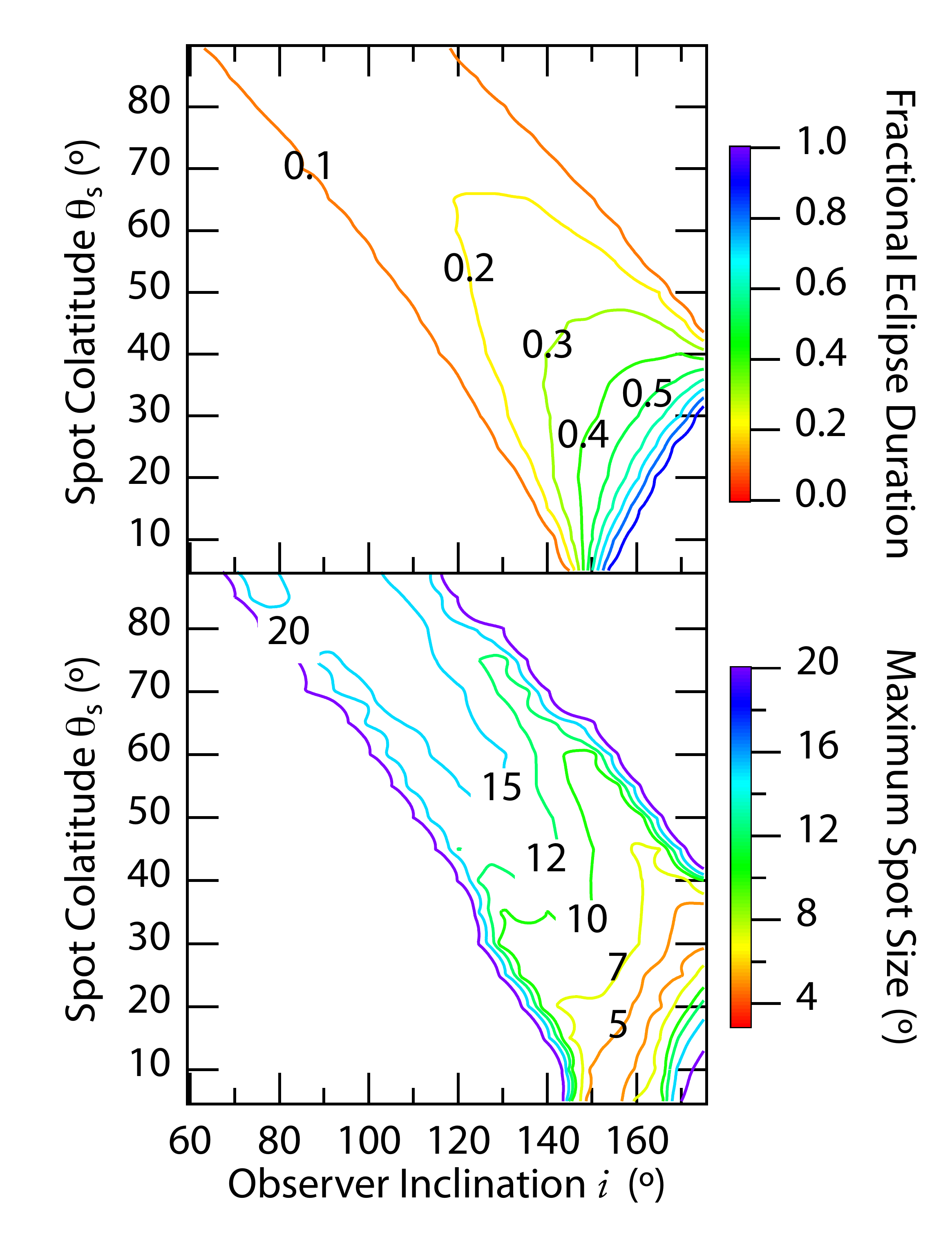, width=3.5in}
\caption{Region of the parameter space in which the spot is eclipsed
  for some fraction of the lightcurve. The upper panel shows the
  fractional duration of the eclipse for a $2^\circ$ spot angular
  radius: zero corresponds to a spot that is visible at all phases,
  while one corresponds to a spot that is never visible. For larger
  spots, the eclipse duration decreases. The lower panel shows the
  spot angular radius at which the duration of the eclipse is shorter
  compared to the eclipse for a 2$^\circ$ spot by 0.1 of the period. }
\label{fig:Max_Spot_Eclipse}
\end{figure}

The upper shaded region in Figure~\ref{fig:Max_Spot_Size} corresponds
to the part of the parameter space where the spot is eclipsed for some
fraction of the spin period. We focus on this region in
Figure~\ref{fig:Max_Spot_Eclipse}. The contours in the top panel of
Figure~\ref{fig:Max_Spot_Eclipse} show the fraction of the profile
during which the spot is eclipsed. Because of the strong lensing, the
spot is hidden from view entirely for only a small portion of the
parameter space, when the spot is near one pole and the observer's
line of sight is near the other.

For the purpose of the top panel in Figure~\ref{fig:Max_Spot_Eclipse},
we have set the spot size to $\rho=2^\circ$. As the spot grows in
size, however, it spends less time completely hidden behind the star
and the eclipse duration decreases. If we were to make the assumption
that the spot was infinitesimal in size, this decrease in the eclipse
duration would introduce a bias in the parameters derived from the
eclipse. In the lower panel of Figure~\ref{fig:Max_Spot_Eclipse}, the contours correspond to 
the spot size that introduces an error equal to 10\% of the total
period to the eclipse duration. Note that the unevenness in the contours is a result of the numerical uncertainty in the eclipse duration. The dominant source of this uncertainty is the phase resolution of the pulse profile, which is on the order of 1\%. As in the case of no eclipses, spot
sizes smaller than 15--20$^\circ$ introduce marginal biases to the
measurements of the eclipse duration and, hence, of the neutron-star
radius.

\section{Sources}

\begin{figure}
\psfig{file=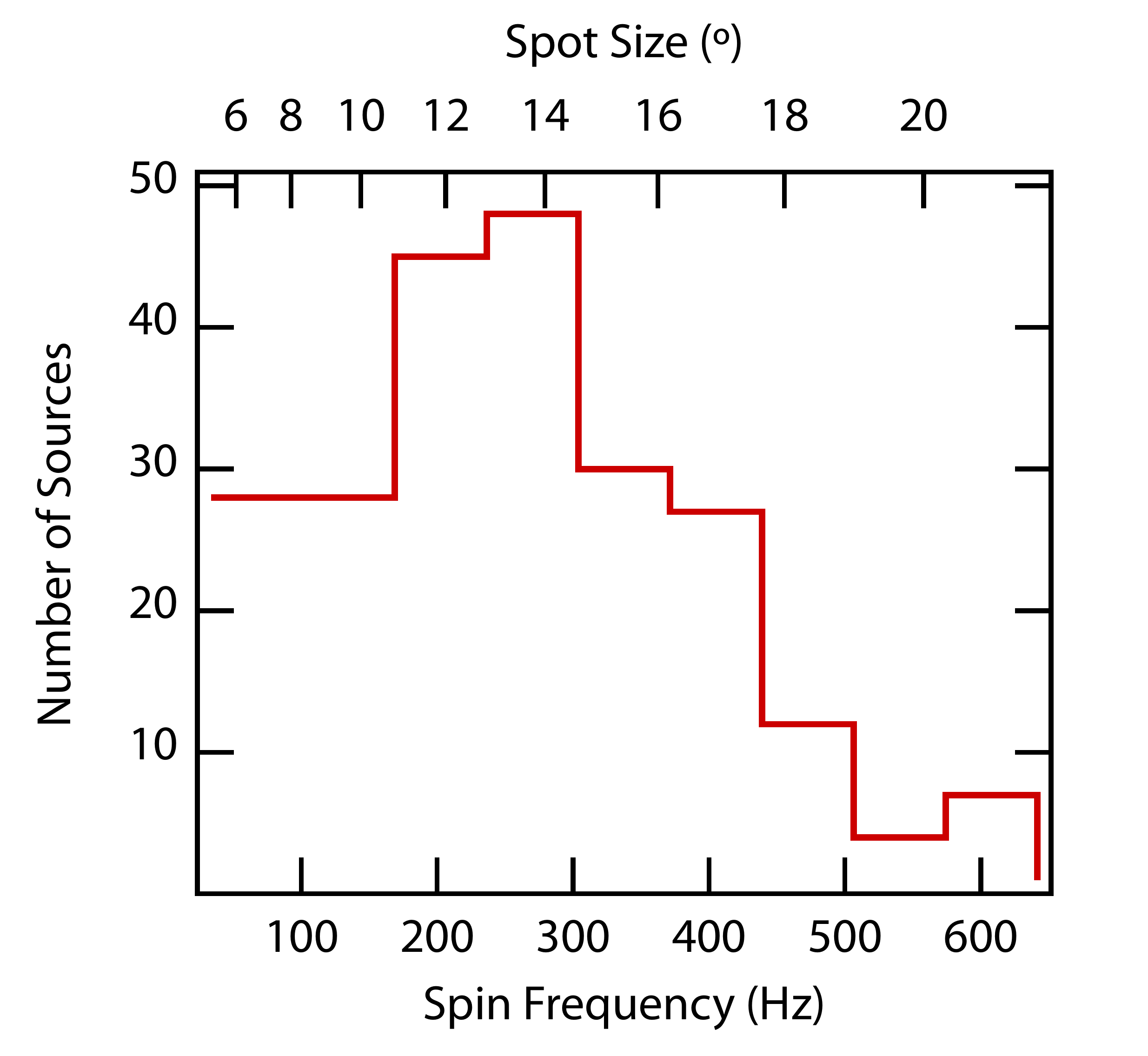, width=3.5in}
\caption{Histogram of the spin frequencies and the corresponding sizes
  of the polar caps of known rotation-powered millisecond pulsars. The
  lower axis shows the spin period, while the upper axis shows the
  spot size corresponding to equation~(\ref{eq:rho_RPP}).}
\label{fig:MSP}
\end{figure}

\begin{figure}
\psfig{file=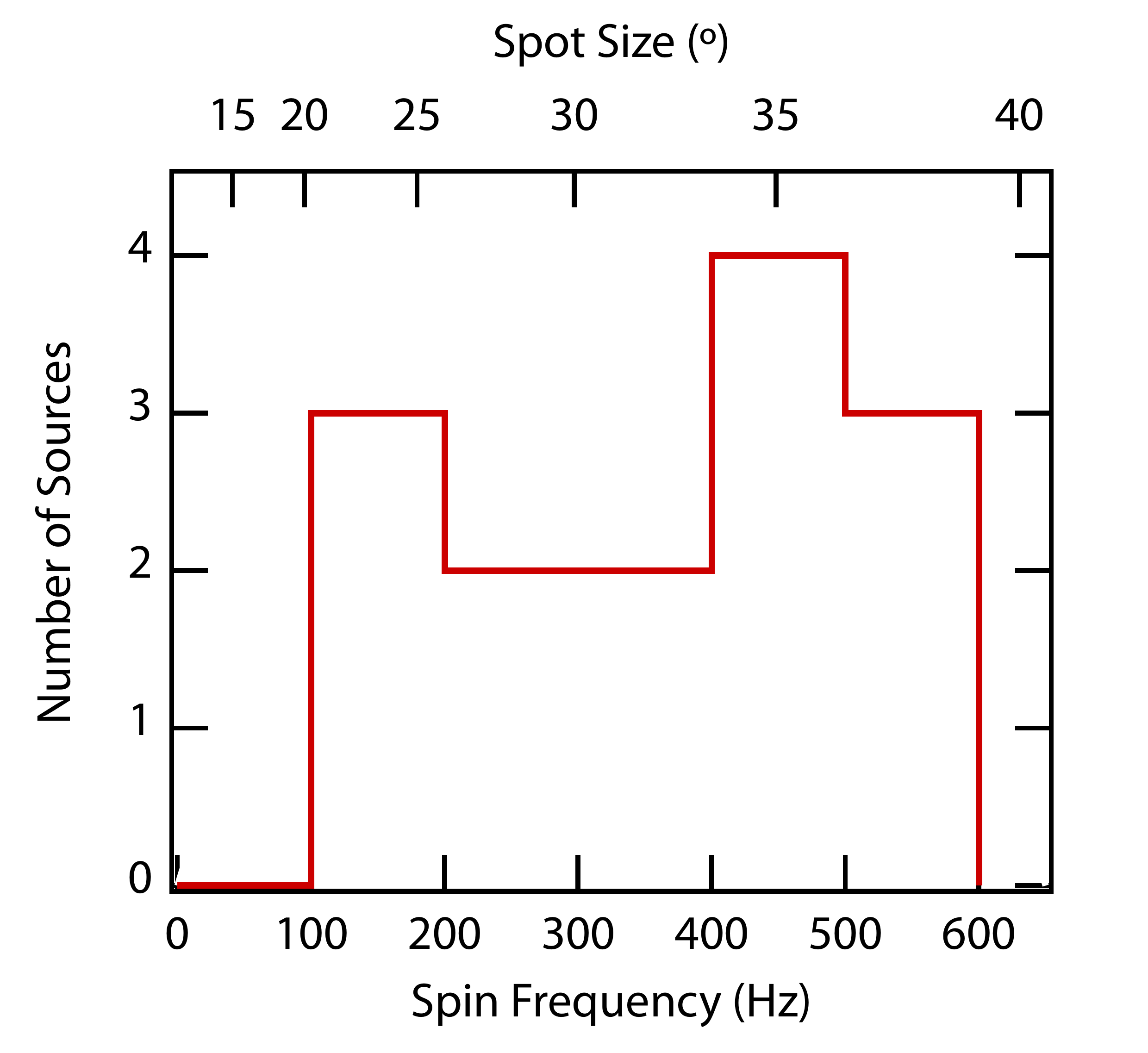, width=3.5in}
\caption{Histogram of accretion-powered X-ray pulsars from Watts
  (2012). As in Figure~\ref{fig:MSP}, the lower axis corresponds to
  the spin frequency, while the upper axis corresponds to the spot
  size derived from equation~(\ref{eq:rho_AMXP}).}
\label{fig:AMXP}
\end{figure}

We now compare our spot size limits to realistic situations where
pulse profiles will be used to measure neutron-star radii with
upcoming missions such as NICER or LOFT. We consider three types of
sources in which X-ray pulsations have been observed. In the case of
rotation-powered and accreting millisecond pulsars, the precise size
of the spot is unknown a priori but can be estimated using physical
arguments. For X-ray bursters, the spot grows during the rise of the
burst until it covers the entire surface of the star. In each case, we
estimate the size of the hotspot depending on the specific parameters
of the system.

In rotation-powered pulsars, hotspots on the surface are generated by
magnetic return currents. To obtain a size estimate, we consider the
simple scenario in which the hotspot corresponds to the footprint of
the open field lines on the stellar surface (Sturrock 1971). Open
field lines are defined as those which pass through the light cylinder
radius $R_{LC}$. This is the radius at which a cylinder corotating
with the neutron star reaches the speed of light, i.e.,
\begin{equation}
R_{LC} \equiv \frac{c}{2 \pi f_{\rm NS}},
\label{eq:RLC}
\end{equation}
where $f_{\rm NS}$ is the spin frequency of the neutron star. In a
dipole magnetic field, the quantity $\sin^2 \theta/r$, where $\theta$
is the polar angle, is constant along field lines, leading to
\begin{equation}
\frac{\sin^2\theta}{R} = \frac{\sin^2\theta_{\rm NS}}{R_{\rm NS}}.
\label{eq:dipole}
\end{equation}
The edge of the hotspot (where $\theta_{\rm NS} = \rho$) corresponds
to the field line that reaches the light cylinder radius at the
equator, where $\theta = 90^\circ$. Therefore, we find the angular
radius of the hotspot to be
\begin{equation}
\rho_{\rm RPP} = \sin^{-1} \left( \sqrt{\frac{2 \pi f_{\rm NS} R_{\rm NS}}{c}} \right).
\label{eq:rho_RPP}
\end{equation}

\begin{figure}
\psfig{file=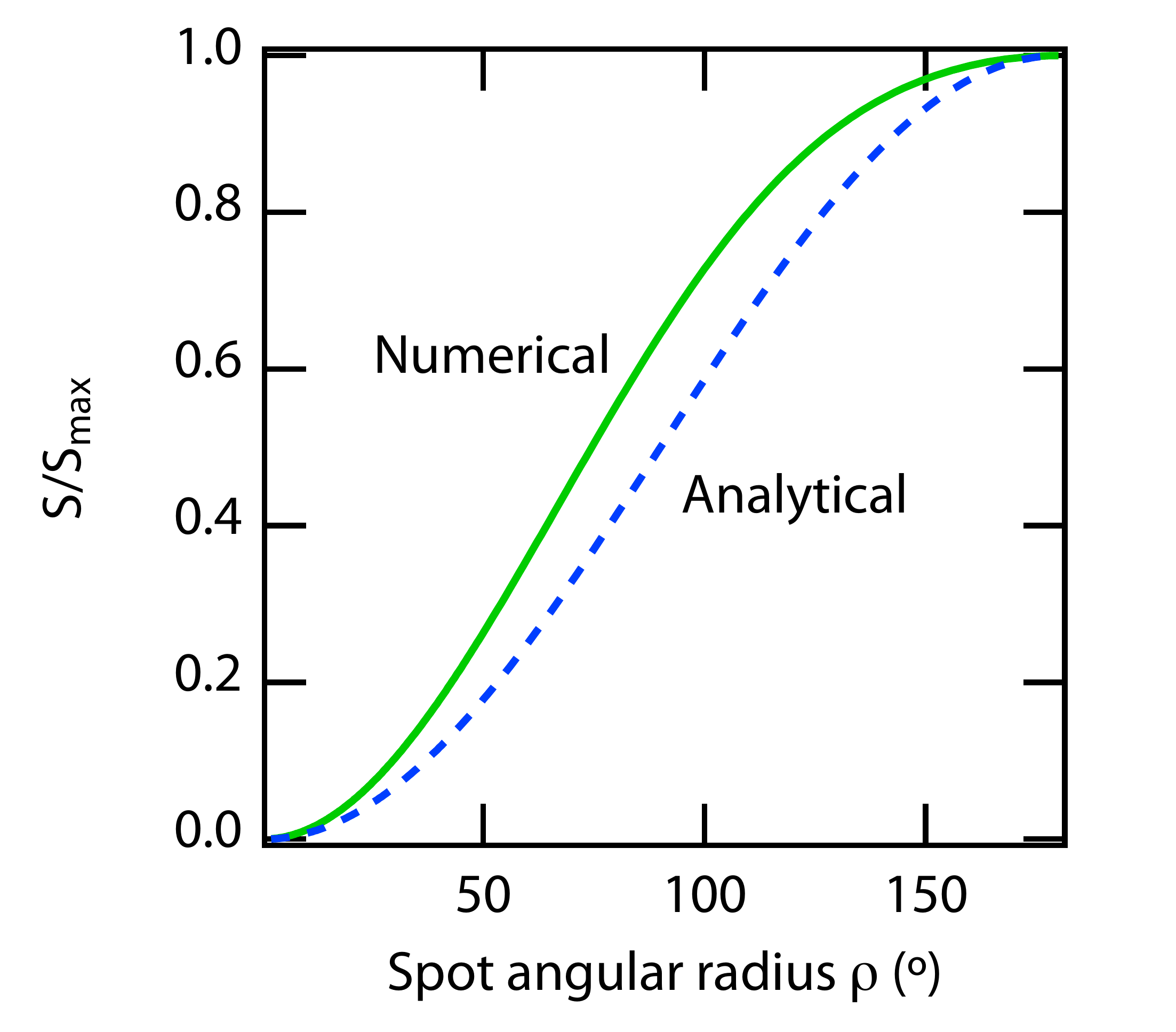, width=3.5in}
\caption{Apparent area of a hotspot averaged over the pulse profile as
  a function of the spot size. The analytical curve corresponds to
  equation~(\ref{eq:A_int}). For the numerical curve, we have chosen a
  spot colatitude and observer inclination of $i = \theta_s =
  45^\circ$. For both curves, we have normalized the result by the
  apparent area of the whole neutron star such that $S(\rho =
  180^\circ)$ is equal to one.}
\label{fig:A_Amax}
\end{figure}

Figure~\ref{fig:MSP} shows the spin the distribution of
rotation-powered X-ray pulsars from the online catalog compiled by D.
Lorimer\footnote{http://astro.phys.wvu.edu/GalacticMSPs/GalacticMSPs.txt}. The
lower axis shows the spin period of the pulsar, while the upper axis
shows the spot size from equation~(\ref{eq:rho_RPP}), assuming a 10~km
neutron star. The peak of the distribution is near 300~Hz,
corresponding to a spot size of $14^\circ$, and the majority of
sources have spot sizes smaller than $18^\circ$. This implies that
assuming that the hotspot on rotation-powered pulsars is
infinitesimally small is a reasonable approximation for all but the
most rapidly spinning sources.

In order to compare our analytical estimate with more detailed
numerical simulations, we make use of the results of Bai \& Spitkovsky
(2010a, b), who modeled the size and shape of the polar cap region
under more realistic vacuum dipole and force-free magnetosphere
conditions. They found that the spot size can vary slightly from the
value derived in equation~(\ref{eq:rho_RPP}), especially for neutron
stars in which the magnetic pole is misaligned from the spin
axis. Moreover, they found that the spot is not circular in shape. For
spots smaller than the limits we derived above, we expect the
spot shape to be unimportant to the pulse profile. For larger spots,
further investigation is needed to find the effect of non-circular
spots on the shape of the pulse profiles.

\begin{deluxetable*}{lcccccccc}
\tablecolumns{7}
\tablecaption{Source Parameters}
\tablehead{
	\colhead{Name} &
	\colhead{$f_{\rm NS}$\tablenotemark{a}} &
	\colhead{$A$\tablenotemark{b}} &
	\colhead{$t_R$\tablenotemark{c}} &
	\colhead{$r_b$\tablenotemark{d}} &
	\colhead{$f_o$\tablenotemark{e}} &
	\colhead{$N_{\rm osc}$ \tablenotemark{f}} &
	\colhead{$N_{\rm bursts}$ \tablenotemark{g}} &
	\colhead{$t_{10\%}$ \tablenotemark{h}} \\
	\colhead{} &
	\colhead{(Hz)} &
	\colhead{(km/10~kpc)$^2$} &
	\colhead{s} &
	\colhead{h$^{-1}$} &
	\colhead{} &
	\colhead{} &
	\colhead{} &
	\colhead{Msec}
}
\startdata
EXO 0748-646	& 552	&  114.0    &	4.9	&  0.24	 &  0.02  &	38  	&  1900  &  27.1 \\
4U 1608-52	& 620	&  324.6    &	3.0	&  0.07	 &  0.13  &	17  	&  131   &  7.0 \\
4U 1636-53	& 581	&  124.6    &	2.6	&  0.22	 &  0.11  &	59  	&  536   &  8.7 \\
4U 1702-429	& 329	&  164.6    &	1.0	&  0.13	 &  0.26  &	359  	&  1381  &  39.0 \\
4U 1728-34	& 363	&  121.6    &	1.0	&  0.20	 &  0.10  &	400  	&  4000	 &  69.3 \\
KS 1731-260	& 524	&  88.4	    &	1.1	&  0.20	 &  0.37  &	73  	&  197	 &  35.6 \\
\enddata
\tablenotetext{a}{Spin frequency}
\tablenotetext{b}{Blackbody normalization}
\tablenotetext{c}{Average rise time}
\tablenotetext{d}{Burst rate}
\tablenotetext{e}{Fraction of bursts with oscillations in the rise}
\tablenotetext{f}{Number of bursts with oscillations needed to constrain $R_{\rm NS}$ to within 10\%}
\tablenotetext{g}{Total number of bursts needed to constrain $R_{\rm NS}$ to within 10\%}
\tablenotetext{h}{Observing time needed to constrain $R_{\rm NS}$ to within 10\%}
\label{tb:bursters}
\end{deluxetable*}

In accreting sources, matter is funneled from the disk onto the
stellar surface along magnetic field lines. We will approximate the
radius at which matter is transferred onto the magnetic field as the
corotation radius $R_c$, i.e.\ the radius where a test mass in
Keplerian orbit is corotating with the neutron star. In this case, it
is straightforward to calculate the mass loading radius,
\begin{multline}
R_c \equiv \left(\frac{G M}{4 \pi^2 f_{\rm NS}^2} \right)^{1/3} \\
\approx 2.46 \left(\frac{R}{10~{\rm km}} \right)^{-1} \left(\frac{M}{1.6~{\rm M_\odot}} \right)^{1/3} 
\left(\frac{f_{\rm NS}}{600~{\rm Hz}} \right)^{-2/3}.
\label{eq:Rc}
\end{multline}
Using equation~(\ref{eq:dipole}), we can again find the size of the hotspot as
\begin{equation}
\rho_{\rm AMXP} = \sin^{-1} \left( \sqrt{\frac{4 \pi^2 R f_{\rm NS}}{G M}} \right).
\label{eq:rho_AMXP}
\end{equation}

\begin{figure}
\psfig{file=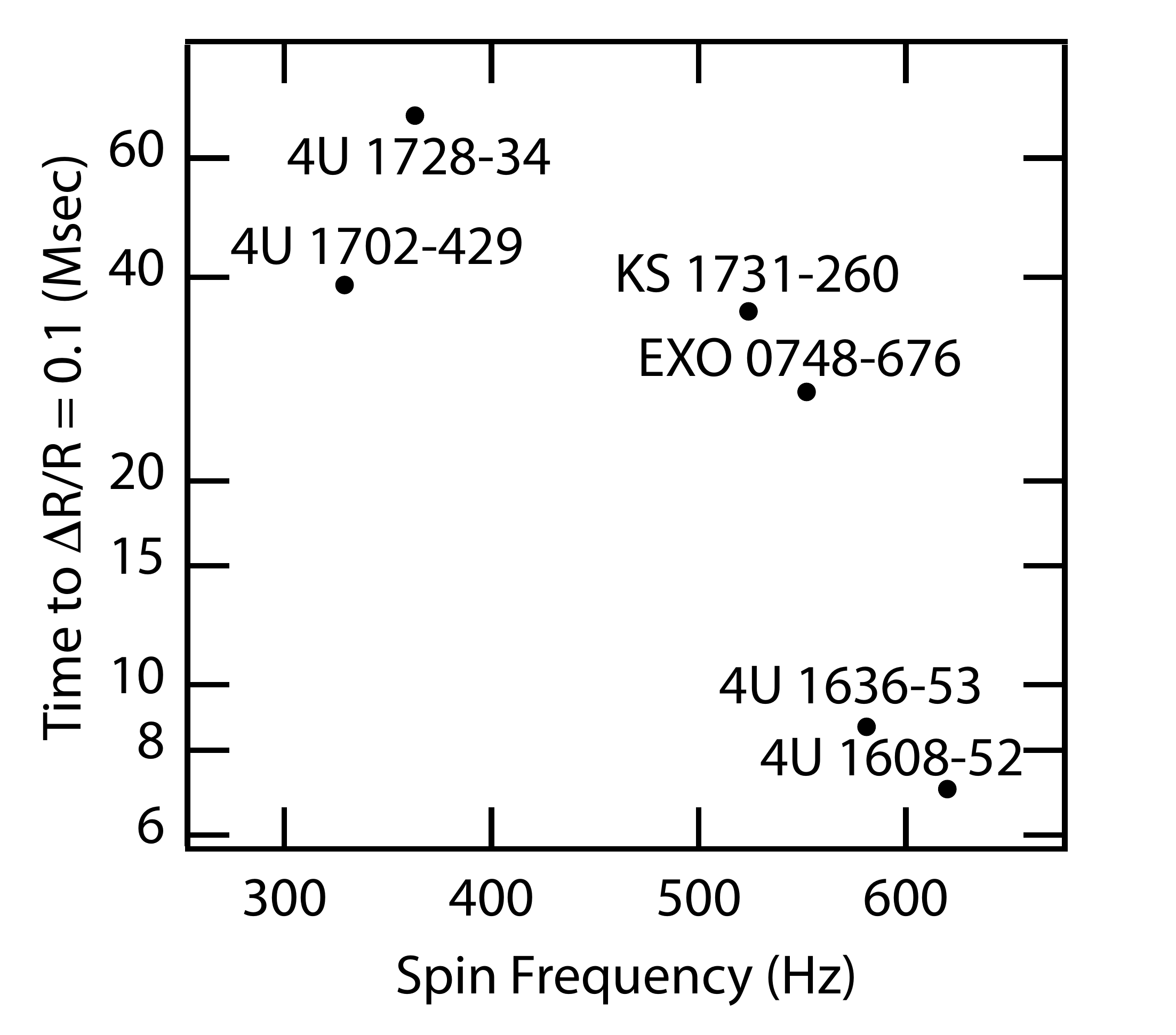, width=3.5in}
\caption{ Time needed to achieve a 10\% accuracy in the measurement of
  neutron-star radii using rise-time burst oscillations for several
  representative X-ray bursters, plotted against their spin
  frequencies.}
\label{fig:Bursters}
\end{figure}

In Figure~\ref{fig:AMXP}, we show the histogram of the spin
frequencies of accreting millisecond pulsars from Watts (2012). As in
Figure~\ref{fig:MSP}, the lower axis shows the spin frequency, while
the upper axis corresponds to the spot size derived from
equation~(\ref{eq:rho_AMXP}). Here, we assume a neutron-star radius of
10~km and a mass of 1.4~M$_\odot$ as before. As Figure~\ref{fig:AMXP}
shows, the majority of accretion-powered pulsars have hotspots that
are significantly larger than the maximum size we calculate
above. Therefore, their pulse profiles will be significantly affected
by the spot sizes and consequently by the spot shapes. Further study
is needed to determine how to extract radius measurements from these
sources.

The last class of sources we consider are X-ray bursters that show
oscillations during the burst rise. In this case, the hotspot on the
surface arises from the relatively slow spreading of the thermonuclear
burning across the stellar surface. Therefore, during the rise, the
spot size changes with time and the amplitude of pulsations decreases
(see Strohmayer et al.\ 1996). We estimate here the fraction of the
burst oscillations for which we can assume that the spot is
infinitesimal without introducing a large bias to the radius
measurement.

Assuming that the spot size increases linearly during the burst rise
from a point at $t = 0$ to a spot covering the entire surface ($\rho =
180^\circ$) at the burst maximum, we write the angular radius as a
function of time,
\begin{equation}
\rho(t) = \frac{\pi t}{t_R},
\label{eq:rhot}
\end{equation}
where $t_R$ is the rise time of the burst. This equation is formally
valid only when the photon diffusion time from the burning layer to
the photosphere is smaller than the lateral propagation time of the
burning front such that the rise time of the burst is dominated by the
lateral spreading. Nevertheless, for the purposes of our estimates
here, we will use this approximation to calculate the time before the
spot reaches some maximum size $\rho_{\rm max}$.

In the Newtonian case, we could calculate the emitting area on the
stellar surface analytically,
\begin{equation}
S_{\rm spot}^{\rm Newton} = \int_0^\rho 2 \pi R_{\rm NS}^2 \sin{\rho'}~d\rho' 
= 2 \pi R_{\rm NS}^2 (1 - \cos{\rho}).
\label{eq:A_int}
\end{equation}
However, this does not take into account the lensing of the spot,
which distorts its size and shape and increases the fraction of the
stellar surface that is visible to a distant observer or the effects
of phase averaging. We numerically calculate the average emitting area
of the spot over all phases as in Psaltis et
al.\ (2000). Figure~\ref{fig:A_Amax} shows the emitting area as a
function of the spot size for a configuration where the spot
colatitude and the inclination to the observer are both
$45^\circ$. The analytic expression derived in
equation~(\ref{eq:A_int}) is shown in the dashed line. As expected in
this configuration, the analytic approximation underestimates the
average flux, since gravitational lensing tends to increase the
apparent surface area.

Converting this surface area into a countrate depends on the specifics
of the spectrum emitted from the hotspot and the detector used. Since
we are primarily interested in an approximate estimate of the time
needed to constrain the radius, we assume that the hotspot emits a
blackbody spectrum at a temperature of 2~keV. The detected flux then
depends only on the angular size of the source, which we encode in the
blackbody normalization $A$, in units of (km/10~kpc)$^2$. The
countrate also depends on the detector efficiency, which we encode in
the quantity $C$ that measures the number of photons detected from a
2~keV blackbody with an angular size of (1~km/10~kpc)$^2$. Then the
countrate that will be observed for a spot of a given size becomes
\begin{equation}
\frac{\rm counts}{\rm second} = \frac{S_{\rm spot}}{S_{\rm max}} A C,
\label{eq:countrate}
\end{equation}
where $S_{\rm max}$ is the apparent area of the whole star. We find
the total number of counts for a given burst by numerically
integrating equation~(\ref{eq:countrate}) from $0$ to $\rho_{\rm
  max}$. In order to estimate total number of photons $N$ for a
typical spot location and observer inclination, we set $\rho_{\rm
  max}$ to $15^\circ$ and C to $2700$~counts~s$^{-1}$~$(10~{\rm
  kpc}/{\rm km})^2$ (for the LOFT Large Area Detector, T. G\"uver 2015,
private communication) and find
\begin{equation}
N = 2.64 A t_R.
\label{eq:count_num}
\end{equation}

The number of counts necessary to measure the radius to 10\% depends
on the magnitude of the Fourier components as well as the geometry of
the hotspot and the spin frequency of the neutron star. Psaltis et
al.\ (2014) found that the number of counts necessary to constrain the
radius to 10\% is approximately
\begin{multline}
N \approx 4.7 \cdot 10^4 \left(\frac{C_1}{0.38}\right)^2 \left(\frac{f_{\rm NS}}{600}\right)^2 
\left(\frac{R_{\rm NS}}{10~{\rm km}}\right)^2\\
\left(\frac{\sin{i}}{0.71}\right)^2 \left(\frac{\sin{\theta_s}}{0.71}\right)^2, 
\label{eq:count_needed}
\end{multline}
which is consistent with numerical simulations by Lo et
al.\ (2013). Here, we have set $i = \theta_s = 45^\circ$ as above. For
this configuration, we also find the first Fourier harmonic $C_1$ to
have a fractional amplitude of 0.38 for a spot size of $\rho =
2^\circ$.

We now combine equation~(\ref{eq:count_num}) with
equation~(\ref{eq:count_needed}) to find the number of individual
bursts which must be added to measure the radius to a 10\%
accuracy. If we further know the burst rate, $r_b$, and the fraction
of bursts showing oscillations during their rise, $f_o$, we can
calculate the observing time necessary for each source as
\begin{equation}
t_{10\%} = \frac{1.79 \cdot 10^4}{A t_R} r_b f_o.
\label{eq:t_tenp}
\end{equation}

We present in Table~\ref{tb:bursters} the parameters for several
bursting sources for which oscillations were detected during the rise
time with the /textit{Rossi X-ray Timing Explorer} (\textit{RXTE}). Here, we use the
burst rise times, the burst frequency, and the fraction of bursts with
oscillations from Galloway et al.\ (2008). We use the blackbody
normalization $A$ from G\"uver et al.\ (2012) for sources 4U~1728-34,
4U~1702-429, KS~1731-260, and 4U~1636-53, from \"Ozel (2006) for
EXO~0748-676, and from G\"uver et al.\ (2010) for 4U~1608-52. We have
chosen only the sources for which the spin frequency and blackbody
normalization are known. In Figure~\ref{fig:Bursters}, we show the
time needed to reach an accuracy of 10\% in measuring the radius
versus the spin frequency for these sources.

It is evident from Figure~\ref{fig:Bursters} that sources with higher
spin rates in general allow for better constraints to the radius. This
is because a faster spin leads to stronger Doppler effects, causing
the pulse profile to become more asymmetric. This results in an
increase in the higher Fourier harmonics and thereby an increase in
the accuracy of the radius measurement.

In the preceding estimate, we used the fraction of bursts with
rise-time oscillations as inferred from \textit{RXTE} observations. Using a
future timing instrument with a larger collecting area such as LOFT
will allow searching for burst oscillations during smaller time
segments in the rise of the bursts. This increased sensitivity will
most likely reveal oscillations in bursts in which the spreading time
is short compared to the typical duration of the \textit{RXTE} time segments or
the location of the ignition of the bursts gives rise to smaller
amplitude oscillations. Those bursts, however, will not add an
appreciable number of counts to the integrated pulse profiles that
will be used for measuring neutron star radii.

\section{Conclusions}
In this paper, we focused on the problem of deriving neutron star
parameters by modeling pulse profiles from hotspots on the stellar
surface. In particular, we addressed an often made assumption that the
spot size can be treated to be infinitesimal in size and investigated
the conditions under which this leads to errors that can be tolerated.
We calculated the maximum size for which the hotspot can be considered
to be infinitesimal, both for derivations of the radius from the
Fourier components of the profile and for profiles which include an
eclipse.

We found that hotspots with angular radius smaller than 10$^\circ$--18$^\circ$
produce profiles that are not significantly different from those with
very small hotspots. The maximum allowed spot size that corresponds to
a 10\% error in the derived radius depends on both the location of the
spot and the inclination of the neutron star spin axis to the
observer's line of sight. In general, increasing the inclination and
decreasing the colatitude lead to a larger allowed spot sizes.

Similarly, we found that the duration of eclipses is relatively
insensitive to the size of the hotspot if it is below
10$^\circ$--20$^\circ$. In this case, increasing observer inclination and
decreasing spot colatitude lead to a smaller maximum spot
size. Nevertheless, for some extreme regions of the parameter space,
even a spot smaller than 5$^\circ$ can have an eclipse duration
that is significantly smaller than that of an infinitesimal spot.

Finally, we considered the implications of spot size limits for the
upcoming NICER and the planned LOFT missions. We showed that, for
rotation-powered millisecond X-ray pulsars, the majority of sources
are expected to have spot sizes smaller than the limits derived
above. For accretion-powered pulsars, we expect only the most slowly
spinning sources to have hotspots small enough that the spot size can
be neglected. For X-ray bursters, we calculated the fraction of the
burst rise-time for which the spot is small enough to constrain the
radius to within 10\%. We found that, for two sources (4U~1636-53 and
4U~1608-52), it is possible to measure the radius to a 10\% accuracy
within a time of 7--9~Ms. Other sources with less optimal
configurations require longer observations of 20--40~Ms in order to
accurately measure their radii.

\acknowledgements

This research was conducted at the University of Arizona and is
supported by NSF grants AST 1108753 and AST 1312034 as well as the
NASA ADP award NNX12AE10G. All ray-tracing calculations were performed
on the El Gato cluster that is funded by NSF grant 1228509.

\end{document}